\documentclass[aps,prd,twocolumn,showpacs,superscriptaddress,nofootinbib]{revtex4}

\usepackage{braket}
\usepackage{amsfonts}
\usepackage{amsmath}
\usepackage{amssymb}
\usepackage{amsthm}
\usepackage{bm}
\usepackage{dcolumn}
\usepackage{epsfig}
\usepackage{graphicx}
\usepackage{graphics}
\usepackage[latin1]{inputenc}
\usepackage{latexsym}
\usepackage{rotating}
\usepackage[colorlinks=true]{hyperref}
\usepackage[usenames]{color}
\usepackage{yfonts}
\usepackage{float}
\usepackage{xspace} % Sensible space treatment at end of simple macros
\usepackage{mathrsfs}
\usepackage{booktabs}
\usepackage[toc,page]{appendix}
\usepackage{siunitx}
\usepackage{array}
\usepackage[normalem]{ulem}
\usepackage{tikz}

\newcommand{\be}{\begin{equation}}
\newcommand{\ee}{\end{equation}}
\newcommand{\EH}{{\mbox{\tiny EH}}}
\newcommand{\CS}{{\mbox{\tiny CS}}}
\newcommand{\EdGB}{{\mbox{\tiny EdGB}}}
\newcommand{\mat}{{\mbox{\tiny mat}}}

\newcommand{\w}{\textbf{l}}
\newcommand{\e}{\textbf{e}}
\newcommand{\n}{\textbf{n}}
\newcommand{\m}{\textbf{m}}

\begin{document}
	\title{Polarization modes of gravitational waves in Quadratic Gravity}
	
	\author{Pratik Wagle}
	\affiliation{eXtreme Gravity Institute, Department of Physics, Montana State University, Bozeman, MT 59717, USA.}
	\affiliation{Department of Physics, University of Illinois at Urbana-Champaign, Urbana, IL 61801, USA.}
	
	\author{Alexander Saffer}
	\affiliation{eXtreme Gravity Institute, Department of Physics, Montana State University, Bozeman, MT 59717, USA.}
	\affiliation{Department of Physics, University of Virginia, Charlottesville, VA 22904, USA.}
	
	\author{Nicol\'as Yunes}
	\affiliation{eXtreme Gravity Institute, Department of Physics, Montana State University, Bozeman, MT 59717, USA.}
	\affiliation{Department of Physics, University of Illinois at Urbana-Champaign, Urbana, IL 61801, USA.}

	\date{\today}
	
\begin{abstract}
	%Abstract character limit is 1300 !!
	%Short abstract below.
	The observation of the inspiral and merger of compact binaries by the LIGO-Virgo collaboration has allowed for new tests of Einstein's theory in the extreme gravity regime, where gravitational interactions are simultaneously strong, non-linear, and dynamical. Theories beyond Einstein's can also be constrained by detecting the polarization modes of gravitational waves. In this paper, we show that dynamical Chern-Simons and Einstein-dilaton-Gauss-Bonnet gravity cannot be differentiated from general relativity based on the detection of polarization modes alone. To prove this result, we use the Newman-Penrose method and an irreducible decomposition method to find that only the tensorial modes can be detected in both these theories. 
		
\end{abstract}

\maketitle

%%%%%%%%%%%%%%%%%%%%%%%%%%%%%%%%%%%%%%%%%%%%%%
\section{Introduction}

General Relativity (GR) has passed a plethora of tests in the Solar System~\cite{Will2014} and in binary pulsars~\cite{Stairs2003}, thus making Einstein's theory one of the most well-verified models in nature. However, these tests have only probed systems in which the gravitational field is either weak, as in the Solar System, or the field is strong but the system is weakly dynamical, as in binary pulsars. Meanwhile, there are some observational and theoretical anomalies that standard GR does not provide a full answer to, such as the late-time acceleration of the universe, the anomalous galaxy rotation curves~\cite{doi:10.1146/annurev.astro.40.060401.093923}, the matter-antimatter asymmetry of the universe, and the existence of singularities. 

A resolution to these anomalies may reside in a modification to Einstein's theory that passes all current tests, yet yields deviations in other extreme regimes, such as where the gravitational interaction is simultaneously strong, non-linear and highly dynamical. On the theoretical side, the intrinsic incompatibility of GR with quantum mechanics has prompted efforts at a variety of unified theories, from string theory~\cite{Polchinski:1998rq,Polchinski:1998rr} to loop quantum gravity~\cite{Alexander:2004xd,Alexander:2009tp,Taveras:2008yf}. On the phenomenological side, the observational anomalies described above have lead to a variety of extensions to GR, such as tensor-vector-scalar theories or TeVeS~\cite{Bardeen:1980kt,Thorne:1980ru,Skordis:2009bf}, modified gravity or MoG~\cite{Brans:1961sx,Sotiriou:2006hs,Clifton:2011jh}, and massive gravity~\cite{deRham:2014zqa} and bigravity~\cite{Blas:2005sz}. Whether any of these attempts at modifying GR has anything to do with reality can only be determined through further experiment and observation. 

A class of theories that have been proposed to resolve some of these anomalies, yet pass current constraints, are those that correct the Einstein-Hilbert action through a scalar field that is non-minimally coupled to squared curvature. One subset of these theories, dynamical Chern-Simons (dCS) gravity, was proposed as a possible way to explain the matter-antimatter asymmetry of the universe by introducing additional parity-violating, gravitational interactions~\cite{Alexander:2004xd,Alexander:2009tp}. Another subset, Einstein-dilaton-Gauss-Bonnet (EdGB) gravity, was proposed to explain late-time acceleration~\cite{Pilo:2011zz,Paulos:2012xe}. Both of these theories can in principle escape current constraints because large deviations from GR are activated only near spacetime singularities~\cite{PhysRevD.98.021503,Yagi:2015oca,Wagle:2018tyk}.

With the observation of gravitational waves (GWs) by the LIGO and Virgo collaboration, it is now possible to probe the highly dynamical and strongly curved, extreme gravity regime~\cite{Rizwana:2016qdq,Berti:2015itd,Will2014,Berti:2018cxi}. The observations of GWs from the mergers of binary compact objects, like black holes and neutron stars, have allowed us to place constraints on a number of modified theories~\cite{Yunes:2016jcc,Nair:2019iur}. GWs are unique and versatile probes to test extreme gravity, as they are weakly interacting, and thus, travel unhindered from their sources to our detectors. The addition of more ground-based and space-based detectors in the near future will lead to numerous observations that will yield the most stringent tests for GR, as well as the most stringent constraints on modified gravity.

In principle, GW observations can also be used to carry out precision studies of their polarizations, particularly allowing for constraints on non-tensorial modes~\cite{Chatziioannou:2012rf,Abbott:2017tlp,Abbott:2018utx}. GR admits only two modes of polarization, i.e.~the $+$ (plus) and $\times$ (cross) polarization modes. A general theory of gravity allows up to six polarization modes; in addition to the two tensorial ones, the other four correspond to two scalar modes (a longitudinal and a ``breathing'' mode) and two vector (transverse) modes. The response of an interferometer depends strongly on the polarization content of the impinging GW. Therefore, if enough detectors receive a signal, one can in principle separate all polarization modes from the data~\cite{Chatziioannou:2012rf}. The presence of only tensorial modes in GW signals could then both verify GR and stringently constrain beyond-Einstein theories that predict additional polarizations. 

One can analytically obtain the polarization modes of a GW in a particular theory by a number of methods. One such method is the study of polarization modes of weak, plane and nearly null GWs using the Newman-Penrose (NP) formalism~\cite{Newman-Penrose}. This method can be employed along with the E(2) classification to calculate NP scalars corresponding to different polarization modes~\cite{Nishizawa:2009bf,Alves:2009eg,Moon:2014qma}. Another method to discover the polarization content of GWs in a given theory is through an irreducible decomposition ~\cite{Flanagan:2005yc,poisson_will_2014}. In this formalism, the metric is reduced into irreducible components, namely a scalar component, two vector component (a longitudinal and a transverse part) and four tensor components (a trace, a longitudinal a trace-free, and a longitudinal-transverse and transverse-tracefree part). Correspondingly, the field equations are reduced into independent scalar, vector and tensorial components, which can be identified with radiative and non-radiative degrees of freedom. 

In this paper, we study EdGB and dCS gravity and calculate its polarization content explicitly. We employ both the NP formalism and the irreducible decomposition method and find that in both theories, GWs possess the two tensorial modes namely, the $+$ and the $\times$ modes just as one would observe for GWs in GR. Therefore, EdGB and dCS gravity are examples of theories for which a polarization test would be completely unconstraining. 

The remainder of this paper deals with the details of the results summarized above. 
Sections~\ref{sec:I} and~\ref{sec:Edgb} provide a basic introduction to dCS and EdGB gravity respectively. 
Sections~\ref{sec:NP} and~\ref{sec:e2}  introduce the NP formalism and the E(2) classification respectively, and we apply it to GR, dCS and EdGB gravity provided in Sects.~\ref{sec:NPGR}, \ref{sec:NPdcs} and \ref{sec:NPedgb} respectively. 
Section \ref{sec:irred} provides a very brief introduction to the irreducible decomposition method, followed by application and analysis for GR, dCS and EdGB in Sects.~\ref{sec:irredgr}, \ref{sec:irreddcs} and \ref{sec:irrededgb} respectively. 
Section \ref{sec:discuss} concludes and points to future research. 

Henceforth, we adopt the following conventions throughout the paper unless otherwise mentioned: we work in 4-dimensions with metric signature $(-,+,+,+)$ as used in ~\cite{Misner:1974qy}, Latin indices (a,b,c,..,j,k,..) in index list represent spatial indices, whereas Greek indices ($\alpha, \beta ....$) represent spacetime indices, round brackets around indices represent symmetrization, $\partial_\mu$ represents a partial derivative, $\Box = \partial_\mu \partial^\mu$ whereas $\nabla^2 = \partial_j \partial^j$, the Einstein summation convention is employed and we work in geometric units in which $G=1=c$.

%%%%%%%%%%%%%%%%%%%%%%%%%%%%%%%%%%%%%%%%%%%%%%
\section{Quadratic gravity theories}
\subsection{ \label{sec:I} Dynamical Chern-Simons Gravity} 

This subsection provides a brief review of dCS gravity and establishes some notation. We will be presenting a minimal review here and direct the interested reader to the recent review paper~\cite{Alexander:2009tp} for a more complete discussion. The action is given by
\be \label{Action1}
S = S_{\EH} + S_{\CS} + S_{\vartheta} + S_{\mat}\,,
\ee
where the Einstein Hilbert term is
\be \label{eq:EH}
S_{\EH} = \kappa \int_{\nu} d^4 x \sqrt{-g} \; R\,,
\ee
with $\kappa = (16 \pi )^{-1}$, $R$ the Ricci scalar and $g$ the determinant of the metric tensor $g_{ab}$.
The CS term is
\be
S_{\CS} =  \frac{\alpha}{4} \int_\nu d^4 x \ \sqrt{-g} \ \vartheta  ~^*\!R R\,,
\ee
where $\alpha$ is a coupling constant, $^*\!R R$ is the Pontryagin density, defined via
\be \label{eq:Pont1}
^*\!R R  \equiv ~^*\! R^{\mu}{}_{\nu}{}^{\kappa \delta} R^\nu{}_{\mu \kappa \delta}\,,
\ee
with $^*\!R^{a}{}_{b}{}^{cd}$ the dual Riemann tensor defined as 
\be
^*\! R^{\mu}{}_{\nu}{}^{\kappa \delta} \equiv \frac{1}{2} \epsilon^\mu{}_{\nu \alpha \beta} ~ R^{\alpha \beta}{}^{\kappa \delta}  \,,
\ee
$\vartheta$ is a pseudo-scalar field and $\epsilon^{abcd}$ is the Levi-Civita tensor. The Pontryagin density can also be expressed as a total divergence of a topological current which contains a combination of the product of Christoffel symbols and its derivatives ~\cite{Wagle:2018tyk}.  The action for the scalar field is
\be \label{eq:sf1}
S_{\vartheta} = - \frac{\beta}{2} \int_\nu d^4 x  \sqrt{-g} \left[ g^{\mu \nu} (\nabla_\mu \vartheta) (\nabla_\nu \vartheta) + 2 V(\vartheta) \right]\,,
\ee
where $\nabla_{\mu}$ is the covariant derivative operator compatible with the metric, $\beta$ is a constant that determines the gravitational strength of the CS scalar field stress energy distribution, while $V(\vartheta)$ is a potential for the scalar that we set to zero. In addition to these terms, one must also include a matter action that couples directly to the metric tensor only.

The field equations for dCS gravity can be obtained by varying the action with respect to the metric tensor and the scalar field. These equations are
\begin{align}
G_{\mu \nu} + \frac{\alpha}{\kappa} & C_{\mu \nu} = \frac{1}{2 \kappa} \left( T^{\mat}_{\mu \nu} + T^{\vartheta}_{\mu \nu} \right)  \,,
\label{eq:field-eq}
\\
\label{eq:CSfield}	\beta \Box & \vartheta + \frac{\alpha}{4} ~^*\! R ~ R = 0 \,,
\end{align}
where $ \Box \equiv \nabla_\alpha \nabla^\alpha$ is the d'Alembertian operator, $T^{\mat}_{\mu \nu}$ is the matter stress energy tensor, $T^{\vartheta}_{\mu \nu} $ is the scalar field stress-energy tensor, $G_{\mu \nu}$ is the Einstein tensor and $C_{ab}$ is the C-tensor, which contains derivatives of the scalar field and the metric and is also trace-free in nature. The stress energy tensor of the scalar field is given by
\be \label{eq:SETfield}
T^{\vartheta}_{\mu \nu} = \beta \left[ (\nabla_\mu \vartheta)(\nabla_\nu \vartheta) - \frac{1}{2} g_{\mu \nu} (\nabla^\sigma \vartheta) (\nabla_\sigma \vartheta) \right]\,.
\ee
The C-tensor in Eq. \eqref{eq:field-eq} can be split into two separate parts, $ C^{\mu \nu} = C^{\mu \nu}_1 + C^{\mu \nu}_2 $, where
\begin{align} \label{eq:CTensor}
C_1^{\mu \nu} & = (\nabla_\sigma \vartheta) \epsilon^{\sigma \delta \alpha (\mu} \nabla_\alpha R^\nu){}_\delta \,,\nonumber \\ 
C_2^{\mu \nu } & = (\nabla_\sigma \nabla_\delta \vartheta) ~ {}^*R^{\delta(\mu \nu) \sigma} \,.
\end{align}

%------------------------------------------------------------------------------------------------------------------------
\subsection{Einstein dilaton Gauss Bonnet Gravity} \label{sec:Edgb}

In this subsection, we provide a brief overview of EdGB gravity. The action in this theory is given by 
\be \label{Action2}
S = S_{\EH} + S_{\EdGB} + S_{\vartheta} + S_{\mat}\,,
\ee
where $S_{\EH}$ and $S_{\vartheta}$ are given by Eq.~\eqref{eq:EH} and Eq.~\eqref{eq:sf1} respectively. The matter action couples only to the metric. The EdGB term is given by
\be \label{eq:edgbterm}
S_{\EdGB} = \int d^4 x \sqrt{-g} \; \lambda \; \vartheta \; \mathcal{G}\,,
\ee
where $\lambda$ is a coupling constant. The Gauss-Bonnet scalar $\mathcal{G} $ can be written in terms of the Riemann tensor as
\be \label{eq:edgbscalar}
\mathcal{G} = \frac{1}{4} \delta^{\mu \nu \alpha \beta}_{\rho \sigma \gamma \delta} R^{\rho \sigma}{}_{\mu \nu} R^{\gamma \delta}_{\alpha \beta} \,,
\ee
with $\delta^{\mu \nu \alpha \beta}_{\rho \sigma \gamma \delta}$ the generalized Kronecker delta. The field equations in EdGB gravity take the form 
\begin{align} \label{eq:EomEdgb}
\nonumber G_{\mu \nu} + 2 \lambda \delta^{\gamma \delta \kappa \epsilon}_{\alpha \beta \rho \sigma} R^{\rho \sigma}{}_{\kappa \epsilon}& (\nabla^\alpha \nabla_\gamma \vartheta) \delta^\beta {}_{(\mu} g_{\nu ) \delta} \\ &= \nabla_{\mu} \vartheta \nabla_\nu \vartheta - \frac{1}{2} g_{\mu \nu} (\nabla_\rho \vartheta \nabla^\rho \vartheta) \,, \\	 \label{eq:EomEdgb2}
\Box \vartheta + &\lambda \mathcal{G} = 0
\end{align}
These equations are obtained by varying the action in Eq.~\eqref{Action2} with respect to the metric $ g_{\mu \nu} $ and the scalar field $ \vartheta $ respectively.

A note of caution regarding notation is now due. It is customary to represent the field that couples to squared curvature with the symbol $\vartheta$ in both dCS gravity and EdGB gravity. However, these fields are not the same. In dCS gravity, $\vartheta$ is a pseudo-scalar field, while in EdGB gravity $\vartheta$ is a scalar field. We will never consider a theory where both the dCS and EdGB corrections to the action are included simultaneously, so it should be straightforward to see what $\vartheta$ represents in any subsequent section of the paper by context.

%%%%%%%%%%%%%%%%%%%%%%%%%%%%%%%%%%%%%%%%%%%%%%
\section{Newman Penrose Formalism} \label{sec:NP}

The study of GWs using tetrad and spinor calculus gained prominence in the 1960s. Ezra Newman and Roger Penrose came up with a formalism that combines these calculus techniques to derive a very compact and useful set of equations that are equivalent to the Einstein equations. This set of equations consists of a linear combination of equations for the Riemann tensor in terms of Ricci rotation coefficients or spinor affine connections ~\cite{Newman-Penrose}. The different possible components of the Riemann tensor or the Weyl tensor in a null tetrad or a null basis were then associated with certain quantities, called Newman-Penrose (NP) coefficients or NP scalars. These coefficients provided physicists with a new tool to understand GWs especially since they relate directly to GW polarization. Later, in ~\cite{Eardley-Lee-Lightman,Eardley-Lee-Lightman-Will}, several authors investigated a formalism to transform from Cartesian coordinates to null tetrads. In this section, we present a brief introduction to the NP formalism and the E(2) classification. We refer the interested reader to~\cite{Will:1993ns} and~\cite{Eardley-Lee-Lightman,Eardley-Lee-Lightman-Will} for a more in-depth discussion.

%------------------------------------------------------------------------------------------------------------------------
\subsection{E(2) classification} \label{sec:e2}

The most general GW that a theory may predict can be composed of six polarization modes in total, which are characterized by the six "electric" components of the Riemann tensor $ R_{0i0j}$, which govern the driving forces in a detector \cite{Eardley-Lee-Lightman}. Indeed, the geodesic deviation equation states that the acceleration of a test particle with spatial coordinates $x^{j}$ with respect to the origin is 
\be
a_{i} = - R_{0i0j} x^j \,,
\ee
where $R_{0i0j}$ are the electric components of the Riemann tensor, due to e.g.~impinging GWs or other external gravitational influences. One can therefore characterize a GW just in terms of the Riemann tensor it produces. 

A weak, plane, nearly null GW in any metric theory can be defined to be a weak, propagating vacuum gravitational field characterized by a linearized Riemann tensor that depends only on the retarded time $\tilde{u}$ i.e.,
\be \label{udep}
R_{\alpha \beta \gamma \delta} = R_{\alpha \beta \gamma \delta}(\tilde{u}) \,,
\ee
with the wave vector normal to the surfaces of constant $ u $, 
\be 
\tilde{l}_\mu= -\tilde{u}_{,\mu}\,.
\ee
This wave vector is \emph{almost} null with respect to a certain local Lorentz metric.
\be
\eta^{\mu \nu} \tilde{l}_\mu \tilde{l}_\nu = \epsilon ~ , ~~~~~~ |\epsilon|\ll 1 \,.
\ee
where $ \epsilon $ is related to the difference in speed as measured in a local Lorentz frame at rest in the universe rest frame, between light and the propagating GW.

Let us now be more formal and begin by introducing a null tetrad as a basis instead of a locally Lorentz orthonormal basis  $(t,x^j)$. For a null plane wave propagating in the $+z$ direction, we define retarded time as $u = t - z$, while if the wave is traveling in the $-z$ direction then advanced time is $v = t + z$. We then define a (completely) null basis $ (l^\mu , n^\mu , m^\mu , \bar{m}^\mu) $ with 
\begin{equation}
l_\mu = -u_{,\mu} ~~ , ~~~~ n_\mu = -\frac{1}{2} v_{,\mu} \,,
\end{equation}
and in the $ (t,x^j) $ basis, our null tetrad vectors can be expressed as
\begin{align}
\nonumber l^\mu &= (1,0,0,1) \\ \nonumber
n^\mu &= \frac{1}{2}(1,0,0,-1) \\
m^\mu &= \frac{1}{\sqrt{2}} (0,1,i,0) \,,
\end{align}
with $ \bar{m}^\mu $ the complex conjugate of $ m^\mu $. These form a null tetrad as each of the individual vectors are orthogonal with respect to themselves, i.e.
\begin{align} \label{eq:ortho}
\nonumber l^\mu l_\mu =&~ 0 = n_\mu n^\mu \\
m^\mu m_\mu =& ~0 = \bar{m}^\mu \bar{m}_\mu \,.
\end{align}
Also, these null vectors obey the orthonormality conditions
\begin{equation} 
-l^\mu n_\mu = m^\mu \bar{m}_\mu = 1
\end{equation}
The Minkowski metric in such a null tetrad can be expressed as 
\be
\eta^{\mu \nu} = - 2 l^{(\mu} n^{\nu)} + 2 m^{(\mu} \bar{m}^{\nu)} \,.
\ee  
which in matrix form is simply
\be \label{eq:minkmat}
\eta^{\mu \nu} = \eta_{\mu \nu}= \begin{bmatrix}
~	0 & 1 & 0 & 0~\\
~	1 & 0& 0& 0~\\
~	0&0&0&-1~\\
~	0&0&-1&0~
\end{bmatrix}
\ee

	\begin{figure}[t] 
	\begin{tikzpicture}[scale=0.50]
	
	\draw [<->,>=stealth] (0,0) -- (0,5);
	\draw [<->,>=stealth] (-3,2.5)--(3,2.5);
	\node (x) at (3.2,2.2) [label=$x$]{};
	\node (y) at (0,5) [label=$y$]{};
	\draw [red,ultra thick](0,2.5) ellipse (0.9cm and 1.8cm);
	\draw [red,densely dotted, ultra thick](0,2.5) ellipse (1.8cm and 0.9cm);
	\node (Rpsi4) at (2.7,0) [label=$ Re ~ \Psi_4$]{};
	\draw (2.1,4.5) ellipse (0.2cm and 0.2cm);
	\draw [ultra thick] (2.1,4.5) ellipse (0.02cm and 0.02cm);
	\node (z) at (2.5,4.3) [label=$z$]{};
	\node (a) at (-2,0) [label=$(a)$]{};
	
	\draw [<->,>=stealth] (8,0) -- (8,5);
	\draw [<->,>=stealth] (5,2.5)--(11,2.5);
	\node (x) at (11.2,2.2) [label=$x$]{};
	\node (y) at (8,5) [label=$y$]{};
	\draw [red,rotate around={45:(8,2.5)}, ultra thick](8,2.5) ellipse (0.9cm and 1.8cm);
	\draw [red,rotate around={45:(8,2.5)},densely dotted, ultra thick](8,2.5) ellipse (1.8cm and 0.9cm);
	\node (Ipsi4) at (10.7,0) [label=$ Im ~ \Psi_4$]{};
	\draw (10.1,4.5) ellipse (0.2cm and 0.2cm);
	\draw [ultra thick] (10.1,4.5) ellipse (0.02cm and 0.02cm);
	\node (z) at (10.5,4.3) [label=$z$]{};
	\node (b) at (6,0) [label=$(b)$]{};
	
	\draw [<->,>=stealth] (0,-7) -- (0,-2);
	\draw [<->,>=stealth] (-3,-4.5)--(3,-4.5);
	\node (x) at (3.2,-4.8) [label=$x$]{};
	\node (y) at (0,-2) [label=$y$]{};
	\draw [blue,ultra thick](0,-4.5) ellipse (1cm and 1cm);
	\draw [blue,densely dotted, ultra thick](0,-4.5) ellipse (1.8cm and 1.8cm);
	\node (phi22) at (2.7,-7) [label=$ \Phi_{22}$]{};
	\draw (2.1,-2.5) ellipse (0.2cm and 0.2cm);
	\draw [ultra thick] (2.1,-2.5) ellipse (0.02cm and 0.02cm);
	\node (z) at (2.4,-2.5) [label=$z$]{};
	\node (c) at (-2,-7) [label=$(c)$]{};
	
	\draw [<->,>=stealth] (8,-7) -- (8,-2);
	\draw [<->,>=stealth] (5,-4.5)--(11,-4.5);
	\node (z) at (11.2,-4.8) [label=$z$]{};
	\node (y) at (8,-2) [label=$y$]{};
	\draw [blue,ultra thick](8,-4.5) ellipse (1.3cm and 1.3cm);
	\draw [blue,densely dotted, ultra thick](8,-4.5) ellipse (1.99cm and 1.3cm);
	\node (psi2) at (10.7,-7) [label=$ \Psi_{2}$]{};
	\draw [ultra thick,->,>=stealth] (9,-2.5)--(10.5,-2.5);
	\node (d) at (6,-7) [label=$(d)$]{};
	
	\draw [<->,>=stealth] (0,-9) -- (0,-14);
	\draw [<->,>=stealth] (-3,-11.5)--(3,-11.5);
	\node (z) at (3.2,-11.8) [label=$z$]{};
	\node (x) at (0,-9) [label=$x$]{};
	\draw [green,rotate around={45:(0,-11.5)},ultra thick](0,-11.5) ellipse (1cm and 1.8cm);
	\draw [green,rotate around={45:(0,-11.5)},densely dotted, ultra thick](0,-11.5) ellipse (1.8cm and 1cm);
	\node (Rpsi3) at (2.7,-14) [label=$ Re~ \Psi_{3}$]{};
	\draw [ultra thick,->,>=stealth] (1.5,-9.5)--(3,-9.5);
	\node (e) at (-2,-14) [label=$(e)$]{};
	
	\draw [<->,>=stealth] (8,-9) -- (8,-14);
	\draw [<->,>=stealth] (5,-11.5)--(11,-11.5);
	\node (z) at (11.2,-11.8) [label=$z$]{};
	\node (y) at (8,-9) [label=$y$]{};
	\draw [green,rotate around={45:(8,-11.5)},densely dotted,ultra thick](8,-11.5) ellipse (1cm and 1.8cm);
	\draw [green,rotate around={45:(8,-11.5)}, ultra thick](8,-11.5) ellipse (1.8cm and 1cm);
	\node (Ipsi3) at (10.7,-14) [label=$ Im~ \Psi_{3}$]{};
	\draw [ultra thick,->,>=stealth] (9,-9.5)--(10.5,-9.5);
	\node (f) at (6,-14) [label=$(f)$]{};
	
	\end{tikzpicture}
	\caption{\label{fig:fig1} (Color online) The impact of the six polarization modes ((a): plus mode, (b): cross mode, (c): breathing mode, (d): longitudinal mode, (e): vector-x mode, (f): vector-y mode) of weak, plane, nearly null GW, permitted in a general, 4-dimensional theory of gravity, on a ring of test particles.  The red, blue and green colors correspond to the tensor, scalar and vector modes respectively. The circled dot in (a), (b) and (c) indicate the wave propagating out of the page. All modes (a)$-$(f) are propagating in the $+z$ direction. The solid line shows the displacement that each mode induces on a ring of test particles in the x$-$y plane, while the dashed line indicates the displacement after half a period.}
\end{figure}
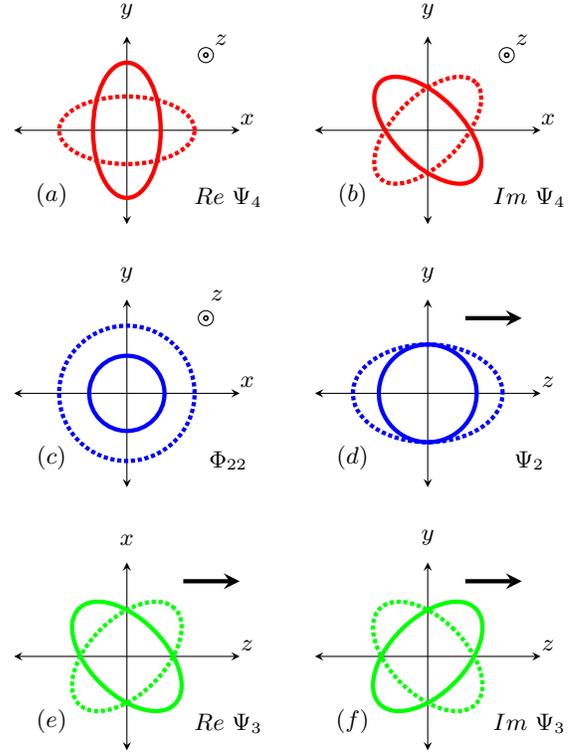

Using such a null tetrad, Newman and Penrose found a set of coefficients (NP coefficients) that describe the radiative modes of the gravitational field~\cite{Newman-Penrose}. These coefficients depend on the Weyl tensor, the traceless Ricci tensor and the Ricci scalar. Correspondingly, these coefficients can also be expressed in terms of the Riemann tensor. Using the geodesic deviation equation and the fact that the Riemann tensor for a GW as defined above just depends on retarded time, it can be shown that the only non-vanishing components of the Riemann tensor are of the form $ R_{npnq} $ with $ p,q \in (l,m,\bar{m})$ in the chosen null tetrad. We have here introduced the contracted tetrad notation, where for example
\be
R_{nlnl} = R_{\mu \nu \alpha \beta} n^{\mu} l^{\nu} n^{\alpha} l^{\beta}\,.
\ee
In general, a tensor in the null basis can be expressed in the Cartesian basis as
\be \label{Trans1}
A_{apb...} = A_{\alpha \beta \gamma ...} a^{\alpha} p^\beta b^\gamma ... ~~ \,,
\ee
where $(a,b,c,\ldots,o)$ can be any of $(l,n,m,\bar{m})$, while $(p,q,r,\ldots,w)$ can only be one of $(l,m,\bar{m})$, while the Greek indices run over $(t,x,y,z)$.

We can also define null vector fields such that $ \w= l^\mu \e_\mu $, $\n = n^\mu \e_\mu$, $\m = m^\mu \e_\mu$ and similar for $\bf{\bar{m}}$. Since these form a complete set of basis vectors, we can expand the GW vector $\tilde{l}^{\mu}$ in terms of them. However, since we are working with a nearly null GW, the expansion of $\tilde{l}^{\mu}$ depends on the velocity of the observer's local frame relative to the global rest frame. Choosing a preferred observer whose frame is at rest with respect to the global rest frame, we have
\be \label{eq:GWwave}
\tilde{l}^\mu = l^\mu (1 + \epsilon_l) + \epsilon_n n^\mu + \epsilon_m m^\mu + \epsilon_{\bar{m}} {\bar{m}}^{\mu} \,.
\ee 
This chosen observer is free to orient her spatial basis such that GWs and her null wave are parallel, and she can choose a frequency such that it is equal to that of GW. These conditions reduce Eq.~\eqref{eq:GWwave} to the form,
\be \label{eq:GWwave2}
\tilde{l}^\mu = l^\mu  - \epsilon_n\left(\frac{1}{2} l^\mu - n^\mu \right)  \,.
\ee 
There are clearly no components of $\tilde{l}^{\mu}$ along $\m^{}$ or \textbf{$\bar{m}^{\mu}$}.

Putting all of this information together, the independent non-vanishing coefficients for a nearly null, plane GW in the preferred tetrad have the form
\begin{align} \label{pol1}
\nonumber \Psi_2 =& -\frac{1}{6} R_{nlnl} \\
\nonumber \Psi_3 =& -\frac{1}{2} R_{nln\bar{m}} \\
\nonumber \Psi_4 =& - R_{n\bar{m}n\bar{m}} \\
\Phi_{22} =& - R_{nmn\bar{m}} \,.
\end{align}
where $ \Psi_3 $ and $ \Psi_4 $ are complex scalars. One can further show that each of these scalars represents a different polarization mode, each of which affects the way matters responds to an impinging GW differently, as shown in Fig.~(\ref{fig:fig1}). 

Let us now consider the functional form NP scalars take for GWs. A GW can be represented in metric form via the decomposition
\be \label{eq:linth}
g_{\mu \nu} = \eta_{\mu \nu} + p_{\mu \nu} \,,
\ee
where $ p_{\mu \nu} $ is the GW metric perturbation. The Riemann tensor for such a linearized metric takes the form
\be \label{linriem}
R_{\mu \nu \alpha \beta} = \frac{1}{2} (p_{\mu \beta , \alpha \nu} - p_{\mu \alpha , \beta \nu} + p_{\nu \alpha , \beta \nu} - p_{\nu \beta , \alpha \mu} ) \,.
\ee
but the Riemann tensor, and thus the GW metric perturbation, must be a function of the retarded time $u$. Therefore, in the null basis, Eq.~\eqref{linriem} can be expressed as
\be \label{linriem1} 
R_{abcd} =\frac{1}{2} (\tilde{l}_c \tilde{l}_b \ddot{p}_{ad} - \tilde{l}_d \tilde{l}_b \ddot{p}_{a c} + \tilde{l}_d \tilde{l}_a \ddot{p}_{b c} - \tilde{l}_c \tilde{l}_a \ddot{p}_{b d}) \,,
\ee
where $(a,b,c,d)$ can again be any of $(l,n,m,\bar{m})$. With this at hand, Eq.~\eqref{pol1} can be rewritten in terms of the corresponding Ricci tensor components or even in terms of the metric perturbation as
\begin{align} \label{pol2}
\nonumber \Psi_{2} &= -\frac{1}{6}~ R_{nl} = \frac{1}{12} ~\ddot{p}_{ll} \,, \\ 
\nonumber \Psi_{3} &= -\frac{1}{2}~ R_{n\bar{m}} = \frac{1}{4}~ \ddot{p}_{l \bar{m}}\,, \\
\nonumber \Psi_4 &= \frac{1}{2} ~\ddot{p}_{\bar{m} \bar{m}} \,, \\
\Phi_{22} &= - \frac{1}{2}~ R_{nn} = \frac{1}{2}~ \ddot{p}_{m \bar{m}} \,.
\end{align}

Based on this, we can now organize gravity theories into different classes. For an exactly null wave, these classes are:
\begin{itemize}
\setlength\parskip{0.0cm}
\setlength\itemsep{0.1cm}
\item Class $II_6$ : $\Psi_{2} \neq 0$. All other NP scalars are observer dependent.
\item Class $III_5$ : $\Psi_{2} = 0, \Psi_{3} \neq 0 $. All other NP scalars are observer dependent.
\item Class $N_3$ : $\Psi_{2} \equiv \Psi_{3} = 0, \Psi_4 \neq 0, \Phi_{22} \neq 0$.
\item Class $N_2$ : $\Psi_2 \equiv \Psi_{3} \equiv \Phi_{22} = 0, \Psi_4 \neq 0 $.
\item Class $O_1$ : $\Psi_2 \equiv \Psi_{3} \equiv \Psi_4 = 0, \Phi_{22} \neq 0 $.
\item Class $O_0$ : $\Psi_2 \equiv \Psi_{3} \equiv \Phi_{22} \equiv \Psi_4 = 0 $.
\end{itemize}
The GWs of GR are therefore of class $N_{2}$, while those of scalar-tensor theories, which contain a breathing mode in addition to the two tensorial modes, are of class $N_{3}$. More details about E(2)- classification can be found in \cite{Will:1993ns}.

%------------------------------------------------------------------------------------------------------------------------
\section{GW polarization through the NP Formalism}

In this section, we present a calculation using the techniques presented in the previous subsection. We begin by presenting a brief calculation to obtain the polarization modes in GR followed by application of the NP formalism to obtain the polarization modes in dCS and EdGB.

\subsection{Polarization modes in GR} \label{sec:NPGR}

Before we start with quadratic theories of gravity like dCS and EdGB, let us first try to calculate the polarization modes using NP scalars in GR as a pedagogical warm-up. We begin by considering the field equation for GR in trace-reversed form. 
\be \label{eq:EOMGR1}
R_{\mu \nu} = 8 \pi (T_{\mu \nu} - \frac{1}{2} g_{\mu \nu} T )\,.
\ee
The right hand side of this equation is zero since we assume GWs are propagating in vacuum, and thus $R_{\mu \nu} = 0$. Therefore, in the chosen null tetrad, we have
\be \label{eq:EOMGR3}
R_{nn} = 0 ~~, R_{nl} = 0 ~~, R_{n\bar{m}} =0 \,,
\ee
and thus, $\Psi_{2} = 0 = \Psi_{3} = \Phi_{22} $. The only unconstrained NP scalar is $\Psi_4$. From this, we conclude that GWs in GR are purely tensorial, i.e.~only the $+$ and $\times$ modes exist, and the theory is of class $N_2$ as already anticipated.

%------------------------------------------------------------------------------------------------------------------------
\subsection{Polarization modes in dCS gravity} \label{sec:NPdcs}

Let us now focus on the polarization modes of GWs in dCS gravity using NP formalism and the E(2) classification formalism discussed in  Sec.~(\ref{sec:e2}).  The essence of this calculation lies in expressing Eq.~\eqref{eq:field-eq} such that we isolate the Ricci tensor on the left-hand side, which we can achieve by reversing the trace. Doing so, Eq.~\eqref{eq:field-eq} becomes
\begin{align} \label{eq:tRricci}
\nonumber R_{\mu \nu} =& \frac{1}{2 \kappa} [T^M_{\mu \nu} - \frac{1}{2} g_{\mu \nu} T^M] + \frac{1}{2 \kappa} [ \beta (\nabla_\mu \vartheta) (\nabla_\nu \vartheta)] \\&~~~~- \frac{\alpha}{\kappa} [(\nabla_\sigma \vartheta) \epsilon^{\sigma \delta \alpha}{}_{(\mu} \nabla_\alpha R_{\nu ) \delta} + (\nabla_\sigma \nabla_\delta \vartheta) {}^*R^\delta{}_{(\mu \nu)} {}^\sigma] \,,
\end{align}
where the first term is a combination of the matter stress energy tensor and its trace, both of which we set to zero henceforth, since again we focus only on GWs propagating in vacuum. The second term is the trace-reversed form of the stress energy tensor of the scalar field in Eq.~\eqref{eq:SETfield}, and the last term is simply the C-tensor in Eq.~\eqref{eq:CTensor} because this quantity is naturally trace free.

Now that we have the first field equation in the form we require, let us analyze the scalar field evolution in Eq.~\eqref{eq:CSfield}. The right-hand side of Eq.~\eqref{eq:CSfield} is the Pontryagin density, but when considering GWs, we must work in the far field limit, where this density vanishes. Thus, we have a second equation of the form,
\be \label{eq:CSFvacuum}
\Box \vartheta = 0\,.
\ee
The above equation simply tells us that the scalar field $\vartheta$ is a free wave, which we can represent as
\be \label{eq:waveeq}
\vartheta = A~ e^{i q^\mu x_\mu } \,,
\ee
where $A$ is its amplitude and $q^\mu$ is its 4-wave number (or familiarly, $(\omega,k_x,k_y,k_z)$ in Cartesian coordinates). Moreover, since the wave operator is that of Minkowski in the far zone, we must have that the scalar wave is null:
\be \label{eq:null}
q^\mu q_\mu = 0 \,.
\ee

With this at hand, and using Eq.~\eqref{eq:waveeq} in Eq.~\eqref{eq:tRricci}, the field equations become
\begin{align} \label{eq:tRicci2}
\nonumber R_{\mu \nu} =&  \frac{\beta}{2 \kappa} [  - ~ A^2 ~ q_\mu q_\nu e^{2 i q\cdot x} ]  \\&~~~~ \nonumber - \frac{\alpha}{\kappa} [(A~ i~ q_\sigma e^{i q \cdot x}) \epsilon^{\sigma \delta \alpha}{}_{(\mu} \partial_\alpha R_{\nu ) \delta}\\&~~~~- (A~q_\sigma q_\delta e^{i q \cdot x}) ~{}^*R^\delta{}_{(\mu \nu)} {}^\sigma] \,,
\end{align}
where $ q \cdot x := q^\mu x_\mu$. The different polarization modes contained in Eq.~\eqref{pol2} can be obtained by considering the individual, independent components of Eq.~\eqref{eq:tRicci2}, which we analyze individually below. 

\subsubsection{Analysis of $ \Psi_{2}$ } \label{sec:psi2}

From Eq.~\eqref{pol2}, we know that
\begin{align} \label{eq:psi2b}
\nonumber 
\Psi_{2} = - \frac{1}{6} R_{nl} =&  -\frac{\beta}{12 \kappa} [  - ~ A^2 ~ q_n q_l e^{2 i q \cdot x} ]  \\&~~~~ \nonumber + \frac{\alpha}{6 \kappa} [(A~ i~ q_\sigma e^{i q \cdot x}) \epsilon^{\sigma \delta \alpha}{}_{(n} \partial_\alpha R_{l ) \delta}\\&~~~~- (A~q_\sigma q_\delta e^{i q \cdot x}) ~{}^*R^\delta{}_{(nl)} {}^\sigma] \,.
\end{align}
Recall that we are considering a weak, plane, nearly null GW, and so the Riemann tensor is only a function of the retarded time (as stated in Eq.~\eqref{udep}). Combining this with Eqs.~\eqref{eq:CSfield}, \eqref{eq:CSFvacuum} and \eqref{eq:waveeq}, we can conclude that the wave vector $q^\mu$ will only have a non-vanishing component along the retarded time, or equivalently along $l^{\mu}$ in terms of the null tetrad under consideration. Thus, the only non-vanishing component is $q^l$ or $q_n$ by means of Eq.~\eqref{eq:minkmat}. This then implies that the first term in Eq.~\eqref{eq:psi2b} does not contribute at all.

Let us now consider the second term of Eq.~\eqref{eq:psi2b}. The Levi-Civita tensor in the second term is non-vanishing only when the superscript indices $\alpha$ and $ \delta$ are equal to $m$ or $\bar{m}$. This is because one of the superscript indices of the Levi-Civita tensor is either $n$ or $l$ (due to the symmetrizer), while the $\sigma$ superscript index must contract onto $q_{\sigma}$, which is non-vanishing only in the $n^{\mu}$ direction. Since the metric perturbation is a function of retarded time only, the Ricci tensor must also be a function of retarded time, which means we can write
\be \label{eq:ricci1}
\partial_\alpha R_{\nu \delta} = \tilde{l}_\alpha \dot{R}_{\nu \delta}  \,,
\ee
and this is the only non-vanishing derivative of the Ricci tensor. By definition of $\tilde{l}^\mu$, we have that $\tilde{l}^\mu = l^\mu$ for a perfectly null GW, whereas for a nearly null GW, we have Eq.~\eqref{eq:GWwave2}. However, the second term of Eq.~\eqref{eq:psi2b} needs the $\alpha$ index to be either $m$ or $\bar{m}$, which means that upon contraction with the $\tilde{l}_{\alpha}$ generated from the partial derivative one finds either $\tilde{l}_m$ or $\tilde{l}_{\bar{m}}$, both of which are zero. Therefore, the second term of Eq.~\eqref{eq:GWwave2} also vanishes. Applying a similar treatment to the third term in Eq.~\eqref{eq:psi2b}, one can easily show that it also vanishes.

With all of this at hand, we then have that $R_{nl} = 0$ and thus 
\be \label{eq:psi2=0}
\Psi_{2} = 0 \,
\ee
in dCS gravity. The physical implication of this mathematical result is that GWs in dCS gravity have no longitudinal modes. 

\subsubsection{Analysis of $\Psi_{3}$}

Let us now follow a similar approach to study $\Psi_{3}$. Equation~\eqref{eq:tRicci2} says that
\begin{align} \label{eq:psi3a}
\nonumber \Psi_{3} =&  -\frac{\beta}{4 \kappa} [  - ~ A^2 ~ q_n q_{\bar{m}} e^{2 i q \cdot x} ]  \\&~~~~ \nonumber + \frac{\alpha}{2 \kappa} [(A~ i~ q_\sigma e^{i q \cdot x}) \epsilon^{\sigma \delta \alpha}{}_{(n} \partial_\alpha R_{\bar{m} ) \delta}\\&~~~~- (A~q_\sigma q_\delta e^{i q \cdot x}) ~{}^*R^\delta{}_{(n \bar{m})} {}^\sigma] \,.
\end{align}
The first term in the above equation vanishes since $q_m = 0$. Following the same arguments as those used for $\Psi_{2}$ one can also show that the second and the third term of Eq.~\eqref{eq:psi3a} vanish, using Eqs.~\eqref{Trans1}), \eqref{linriem1} and the orthogonality conditions in Eq.~\eqref{eq:ortho}. 

Combining these results with Eq.~\eqref{pol2}, we then find
\be \label{eq:psi3b}
\Psi_3 = 0 \,.
\ee
The physical interpretation of this mathematical result is that GWs in dCS gravity have no vector modes.  

\subsubsection{Analysis of $\Phi_{22}$}

Let us now study the breathing mode. Equation~\eqref{eq:tRicci2} says that
\begin{align} \label{eq:phi22a}
\nonumber \Phi_{22}  =&  -\frac{\beta}{4 \kappa} [  - ~ A^2 ~ q_n q_{n} e^{2 i q \cdot x} ]  \\&~~~~ \nonumber + \frac{\alpha}{2 \kappa} [(A~ i~ q_\sigma e^{i q \cdot x}) \epsilon^{\sigma \delta \alpha}{}_{(n} \partial_\alpha R_{n ) \delta}\\&~~~~- (A~q_\sigma q_\delta e^{i q \cdot x}) ~{}^*R^\delta{}_{(n n)} {}^\sigma] \,.
\end{align}
As before, the second term in the above equation vanishes by arguments similar to those presented in Sec. \ref{sec:psi2}, whereas the third term vanishes by the definition of the wave 4-vector, the dual Riemann tensor and the Levi-Civita tensor. However, the first term does not vanish by the characteristics of the GW established previously.

The above arguments imply that, in general, $\Phi_{22}$, and thus the breathing mode of GWs in dCS gravity is not vanishing. However, GWs are always defined in terms of the $1/r$ part of the radiative field. Since $ q_n $ falls off as $\mathcal{O}(r^{-1})$, it is then clear that $ \Phi_{22}$ falls of as ${\cal{O}}(r^{-2})$. In the far field, then, we have that
\be \label{eq:phi22b}
\Phi_{22} \rightarrow 0 ~~~~ as ~~~~ r \rightarrow \infty
\ee
and the breathing mode of GWs in dCS gravity vanishes.

\subsubsection{Analysis of $\Psi_4$}

Combining Eq.~\eqref{pol1}, \eqref{pol2} and \eqref{eq:tRicci2}, it can be seen that there are no constraints possible on the $\Psi_4$ mode. Thus, in dCS gravity, $\Psi_4$, or the $+$ and $\times$ polarization modes, cannot be constrained with the use of the field equations.

\vspace{1cm}

From the above analysis, we can see that $\Psi_{4}$ and $\Phi_{22}$ are the non-vanishing NP scalars in general with $\Phi_{22}$ having a $\mathcal{O}(r^{-2})$ dependence. Therefore, one might be able to conclude that dCS is a Class $N_3$ theory behaving like a Class $N_2$ theory in the far field limit. However, this is not entirely true. The reason lies in the definitions of Eq.~\eqref{pol2}. GWs in general are defined as the $1/r$ part of the radiative field far away from the source. This means that we only need to take into consideration the terms of Eq.~\eqref{eq:tRicci2} that are non-vanishing and scale as $\mathcal{O}(r^{-1})$. We can therefore conclude that the non-vanishing part of $R_{nn}$ or $\Phi_{22}$ is not the term that actively contributes to the GW. With all these arguments, we can conclude that for a weak, plane, nearly null GW, an observer can only detect the $+$ and $\times$ polarization modes, just as one would observe in GR. Under the E(2)-classification, dCS therefore always behaves like a Class $N_2$ theory.

At this junction, it is convenient to compare the above result to that of scalar-tensor theories. In such theories, the field equations are quite similar to those in Eq.~\eqref{eq:tRricci}, except for three observations. First, there is no non-minimal coupling between the scalar field and quadratic curvature invariants, so the third terms on the right-hand side of Eq.~\eqref{eq:tRricci} is absent. Second, the second term on the right-hand side of this equation is multiplied by $\vartheta^{-2}$, but since the field is typically assumed to have some cosmological boundary value, this term is still quadratic in the amplitude of the scalar field perturbation. Third, the field equations of scalar-tensor theories have an extra term on the right-hand side of Eq.~\eqref{eq:tRricci} that is proportional to $\vartheta^{-1} \partial_{\mu \nu} \vartheta$, which is linear in the amplitude of the scalar field perturbation. This term arises because the scalar-tensor action has a scalar field multiplying the Ricci scalar, which then leads to non-vanishing contributions when varying the action with respect to the metric tensor and integrating by parts. It is this term in the action of scalar-tensor theories that generates a non-vanishing $\Phi_{22}$ NP scalar, and thus a non-vanishing breathing mode. In dCS, however, this extra term is not present because the Ricci scalar is not multiplied by $\vartheta$, with the field only coupling to the metric through squared curvature invariants, thus explaining why dCS gravity does not possess a breathing mode.

%------------------------------------------------------------------------------------------------------------------------
\subsection{Polarization modes in EdGB gravity} \label{sec:NPedgb}

Let us finally consider the polarization modes of GWs in EdGB gravity. We will not provide here as many details as in the dCS gravity case, as the procedure is fairly similar in essence. Instead, we refer the interested reader to Appendix \ref{Appendix A}.

The evolution equation for the scalar field is identical to that in dCS gravity. This is because the geometry of the source of our GWs does not affect the observer, so once more, one finds a wave equation in flat spacetime for the scalar field, whose solution is of the form of Eq.~\eqref{eq:waveeq}. The evolution of the GW metric perturbation is controlled by the trace-reversed form of field equations in Eq.~\eqref{eq:EomEdgb}. Such a trace-reversed form is 
\begin{align} \label{eq:tricci3}
\nonumber R_{\mu \nu} =&   [T^M_{\mu \nu} - \frac{1}{2} g_{\mu \nu} T^M] +  [  (\partial_\mu \vartheta) (\partial_\nu \vartheta)] \\&~~~~- 2 \lambda~ \delta^{\gamma \delta \kappa \epsilon}_{\alpha \beta \rho \sigma} R^{\rho \sigma}{}_{\kappa \epsilon} (\partial^\alpha \partial_\gamma \vartheta) \delta^\beta {}_{(\mu} g_{\nu ) \delta} \,.
\end{align}
which clearly takes a form quite similar to Eq.~\eqref{eq:CSfield}. Again, since we are working in vacuum, the first term of the above equation is zero. The second term is the same as that in dCS gravity, and thus, it is formally non-vanishing only for the breathing mode, but then again it vanishes in the far field limit. The third term also vanishes for a plane, null GW propagating along the null direction associated with retarded time. This can be shown through Eqs.~\eqref{eq:minkmat}, \eqref{eq:waveeq} and \eqref{linriem1}, along with the orthogonality conditions in Eq.~\eqref{eq:ortho}. Just like in the case for dCS, the $\Psi_4$ mode remains unconstrained.

These arguments lead us to the conclusion that GWs emitted in EdGB gravity can only possess GW polarization modes associated with $\Psi_4$ and $\Phi_{22}$. However, as discussed earlier, $\Phi_{22}$ does not contribute to the GW perturbation, because GWs are defined as the $1/r$ part of the radiative perturbation. Therefore, the only true non-vanishing NP scalar is $\Psi_{4}$. This means that, just like for GR, only the $+$ and $\times$ polarization modes are non-vanishing, thereby making EdGB a class $N_2$ theory under the E(2)-classification.

%%%%%%%%%%%%%%%%%%%%%%%%%%%%%%%%%%%%%%%%%%%%%%
\section{Irreducible decomposition} \label{sec:irred}

In this section, we present a brief overview of an alternate way of identifying the polarization modes by decomposing the metric into irreducible components. Such a decomposition allows us to clearly identify the degrees of freedom present in any theory. As far as we know, this treatment for a linearized theory was first suggested in~\cite{Flanagan:2005yc}.

The metric perturbation transforms as a tensor field under Lorentz transformations in Minkowski spacetime. Such a transformation includes boosts and rotations. Ignoring the boosts and just focusing on pure rotations, $ p_{00} $ transforms as a scalar, $p_{0j} $ transforms as a 3-vector and $ p_{jk} $ transforms as a 3-tensor, where $p_{\mu \nu}$ is the metric perturbation as defined in Eq.~\eqref{eq:linth}. In the Cartesian basis, one can think of the $(0,0)$ component of the metric perturbation as the $(t,t)$ component, while the subscripts $(j,k) \in (x,y,z)$.   These quantities can be decomposed further into their irreducible pieces. Thus, $ p_{0j} $ decomposes into a longitudinal and a transverse piece, whereas $p_{jk} $ decomposes into a trace, a longitudinal and trace-free piece, longitudinal and transverse piece and a transverse and trace-free piece. Thus, such a decomposition portrays all possible degrees of freedom contained in the metric perturbation $ p_{\mu \nu} $.

Following~\cite{poisson_will_2014}, one can express these components of the metric perturbation as
\begin{align} \label{decomp1}
\nonumber p_{00} =~ &2 ~ U \\
\nonumber p_{0j} = ~- 4& ~ U_j - \partial_j A \\
p_{jk} =~ 2 \delta_{jk} V + (\partial_{jk} - \frac{1}{3} \delta_{jk} \nabla^2)& B + (\partial_j B_k + \partial_k B_j) + h_{jk}^{TT} \,,
\end{align}
where $U,V,A,B$ are scalars, $ U_j $ and $B_j$ are 3-vectors and $h_{jk}^{TT}$ is a transverse-traceless 3-tensor. These quantities satisfy the conditions
\begin{align} \label{cond1}
\nonumber \partial_j U^j =& 0 \\
\nonumber \partial_j B^j =& 0 \\
\partial_k h^{jk}_{TT} = 0 = &\delta_{jk} h^{jk}_{TT} \,,
\end{align}
which imply that $U^{j}$ and $B^{j}$ are transverse 3-vectors.

All these individual pieces of the metric perturbation are also gauge invariant. Thus, without loss of generality, we can choose the Coulomb gauge, in which
\be \label{coulomb}
A ~ = ~ B~ = ~ B_j ~ =~ 0 \,,
\ee
such that the components of metric perturbation become
\begin{align} \label{decomp2}
\nonumber  p_{00} =&~ 2 ~ U \\
\nonumber p_{0j} =& ~-4  U_j  \\
p_{jk} =~ 2 \delta_{jk}& V + h_{jk}^{TT} \,,
\end{align}
From these components, we can also construct gauge-invariant potentials, which are given by
\be \label{pots}
\Phi = U, ~~~~~~ \Phi_j = U_j, ~~~~~~ \Psi=V \,,
\ee
and which are clearly essentially equivalent to the gravitational potentials in the Coulomb gauge, making this gauge meaningful and convenient. The gauge invariant potentials represent the degrees of freedom of the gravitational field. The two scalar potentials, $\Phi$ and $\Psi$, the 3-vector potential $\Phi_j$ with its two independent components, and the 3-tensor potential $h_{jk}^{TT}$ with its two independent components, represent the six independent degrees of freedom required to describe all possible GW polarizations in a generic theory of gravity (see also Fig.~(\ref{fig:fig1})).

%------------------------------------------------------------------------------------------------------------------------
\subsection{Irreducible decomposition in GR} \label{sec:irredgr}

We can now apply the treatment mentioned above to reduce the field equation of GR into independent irreducible pieces. For a GW, the linearized Einstein tensor depends on the metric perturbation via
\begin{align} \label{Eins}
\nonumber G_{\alpha \beta} = - \frac{1}{2}(\Box p_{\alpha \beta}+\partial_{\alpha \beta}& p  - \partial_{\alpha \mu} p^\mu{}_\beta - \partial_{\beta \mu} p^\mu{}_\alpha) \\ &+\frac{1}{2} \eta_{\alpha \beta}(\Box p - \partial_{\mu \nu} p^{\mu \nu})\,,
\end{align}
where $ p = \eta_{\alpha \beta} p^{\alpha \beta}$ is the flat spacetime trace. Using Eqs.~\eqref{decomp2},~\eqref{pots}~and~\eqref{Eins}, we can express the individual components of the Einstein tensor as
\begin{align}\label{Einsdec}
\nonumber G_{00} = - 2& \nabla^2 \Psi \\ \nonumber \\
\nonumber G_{0j} = - 2 \partial_{tj}& \Psi + 2 \nabla^2 \Phi_j \\ \nonumber \\
\nonumber G_{jk} = - \delta_{jk}\nabla^2(\Phi-\Psi) &- 2 \delta_{jk} \partial_{tt} \Psi + \partial_{jk}(\Phi-\Psi) \\  ~~~~~~+ 2(\partial_{tj}\Phi_k + \partial_{tk}&\Phi_j) - \frac{1}{2} \Box h_{jk}^{TT}		\,.
\end{align}
The Einstein tensor is now fully decomposed into its irreducible pieces. Now, the right hand side of the field equation, Eq.~\eqref{eq:EOMGR1} involves a stress energy tensor. 

Let us now focus on the right-hand side of the Einstein equations. The stress energy tensor can be decomposed into its own irreducible pieces as
\begin{align} \label{stress1}
\nonumber T^{00} &= \rho \\
\nonumber T^{0j} = (s^j &+ \partial^j s) \\
T^{jk} = \tau \delta^{jk} + \partial^{jk} \sigma - \frac{1}{3} \delta^{jk}& \nabla^2 \sigma + 2 \partial^{(j} \sigma^{k)} + \sigma^{jk} \,,
\end{align}
where $\rho$ is the mass density of the matter distribution measured by an observer at rest, $( s^j + \partial^j s)$ or $T^{0j}$ is the momentum density, and $T^{jk}$ is the stress tensor. These quantities satisfy the conditions
\be \label{Tcond}
\partial_j s^j = 0, ~~~~~~ \partial_j \sigma^j = 0, ~~~~~~ \partial_k \sigma^{jk} = 0 = \delta_{jk} \sigma^{jk} \,,
\ee
which imply that $s^{j}$ and $\sigma^{jk}$ are transverse. 

Energy-momentum conservation in linearized theory reveals that not all of the ten fields in Eq.~\eqref{stress1} are independent. Using that $\partial_\beta T^{\alpha \beta} = 0$, one finds that
\begin{align}\label{eq:redun}
\nonumber \nabla^2 s &= - \partial_t \rho \\
\nonumber \nabla^2 \sigma^j &= - \partial_t s^j \\
\nabla^2 \sigma &= - \frac{3}{2} (\partial_t s + \tau) \,.
\end{align}
This implies that only $\rho$, $s^j$, $\tau$ and $\sigma_{jk}$ are independent, while the other four fields can be determined in terms of them through the equations above. Combining these expressions with the linearized Einstein equations implies that
\begin{align} \label{FE1}
\nonumber \nabla^2 \Psi &= - 4 \pi \rho \,, \\
\nonumber \nabla^2 (\Phi - \Psi) &= -12 \pi (\partial_t s + \tau) \,, \\
\nonumber \nabla^2 \Phi_j &= -4 \pi s_j \,, \\
\Box h_{jk}^{TT} &= -16 \pi \sigma_{jk} \,.
\end{align}
The first three equations above are (elliptic) Poisson equations, and therefore, the solutions at a particular time depend only on the matter configuration at that particular time. In this sense, the 4 degrees of freedom contained in $(\Psi, \Phi, \Phi_{j})$ are constrained by the field equations and do not represent radiative modes. On the other hand, the last equation is a (hyperbolic) wave equation in flat spacetime, which means that $h_{jk}^{TT}$ clearly represents a radiative mode, corresponding to the two polarization modes $(h_+, h_\times)$ of GR. Such a gauge invariant formulation of GWs thus separates the radiative modes from the non-radiative ones. 

%------------------------------------------------------------------------------------------------------------------------
\subsection{Irreducible decomposition in dCS gravity} \label{sec:irreddcs}

Let us now focus on dCS gravity and compare the results of an irreducible decomposition to those obtained from the NP method in Sec.(\ref{sec:NPdcs}).

Consider the field equations in dCS gravity in Eq.~\eqref{eq:field-eq}). The left-hand side of this equation is completely geometric in nature, whereas the right hand side depends on the matter-energy and scalar field content of the system under consideration. We can again decompose our metric perturbation into scalar, vector and tensor parts, as in Eq.~\eqref{decomp2}, which allows us to decompose the geometric part of Eq.~\eqref{eq:field-eq}. The right-hand side of the field equations consists of two independent parts -- matter stress-energy tensor and a scalar field stress-energy tensor --, both of which we can again decompose into irreducible pieces using Eqs.~\eqref{stress1}, \eqref{eq:SETfield} and \eqref{decomp2}. Since Eq.~\eqref{eq:CSfield} behaves like a wave equation in the far zone, the scalar field can be expressed by means of Eq.~\eqref{eq:waveeq} as before.

By following the steps mentioned above, we obtain the decomposed field equations for dCS gravity
\allowdisplaybreaks[4]
\begin{widetext}
\begin{align} \label{eq:FES2}
 \nabla^2 \Psi + \frac{ \alpha}{\kappa} \epsilon_{\alpha \gamma \delta \eta} n^\alpha n^\beta q^\gamma \vartheta (\partial^\eta {}^\delta {}_\beta \Psi) =  \frac{\beta}{8 \kappa} q_\mu q^\mu \vartheta^2 + &\frac{\beta}{4 \kappa} n^\alpha n^\beta q_\alpha q_\beta  \vartheta^2 - \frac{\rho}{4 \kappa} \\ \nonumber \\ 
 \label{eq:FEV2}
\nonumber 2 \nabla^2 \Phi_j - \frac{4 \alpha}{\kappa} \epsilon_{j \beta \gamma}{}^{ \delta} q^\alpha q^\beta \vartheta \partial_\alpha {}_\delta \Phi^\gamma + \frac{2 \alpha}{\kappa} \epsilon_{j \beta \delta \eta} n^\beta n^\alpha q^\gamma q^\delta \vartheta \partial_\alpha {}^\eta \Phi_\gamma  &- \frac{2 \alpha}{\kappa} \epsilon_{\beta \gamma \delta \eta} n^\beta n^\alpha q^\gamma \vartheta \partial^\eta{}_\alpha{}^{(\delta|} \Phi_{|j)} \\  - \frac{\alpha}{\kappa} \epsilon_{j \gamma \delta \eta} n^\alpha n^\beta q^\gamma \vartheta \partial^\eta{}_{\alpha \beta} \Phi^\delta - \frac{2 \alpha}{\kappa} \epsilon_{\beta \gamma \delta \eta} n^\alpha n^\beta q_\alpha q^\gamma \vartheta \partial_j{}^\eta \Phi^\delta &= \frac{\beta}{2 \kappa} n^\alpha q_\alpha q_j \vartheta^2 - \frac{s_j}{2 \kappa} \\ 
\nonumber \\ 
\label{eq:FES3}
\delta_{jk} \nabla^2 (\Phi - \Psi) - \frac{\alpha}{2 \kappa} \epsilon_{(k|\alpha \beta \gamma} q^\alpha \vartheta \partial^{\gamma \beta}{}_{|j)}(\Phi - \Psi)  =&~ - \frac{3}{4 \kappa} \delta_{jk} (\partial_t s + \tau) + \frac{\beta}{2 \kappa} q_j q_k \vartheta^2 \\
\nonumber \\ \label{eq:FET2}
\Box h^{TT}_{jk} + \frac{4 \alpha }{\kappa} [\epsilon_{(k|\beta}{}^{\gamma \delta} q^\alpha q^\beta \vartheta \partial_{\alpha \delta}h^{TT}_{|j)\gamma}     + \frac{1}{2} \epsilon_{(k|\alpha}{}^{\beta \delta} q^\alpha  \vartheta \partial_{\delta} \Box h^{TT}_{|j)\beta} & -  \epsilon_{(k|\beta}{}^{\gamma \delta} q^\alpha q^\beta \vartheta \partial_{|j)\delta}h^{TT}_{\alpha \gamma}] = -\frac{\sigma_{jk}}{\kappa} \,,
\end{align}
\end{widetext}
where we have used the notation $\partial_{\alpha \beta} = \partial_\alpha \partial_\beta$ and $ \partial_{\alpha \beta \gamma} = \partial_\alpha \partial_\beta \partial_\gamma $, and where $n^\alpha $ is the normal 4-vector pointing along the direction of propagation. 
These equations are analogous to Eq.~\eqref{FE1}, with certain modifications that depend on the scalar field, so let's analyze them term by term.

Before doing so, however, it is useful to remember a few facts we discovered in the NP method section. For the GWs under consideration, we have already established that the Riemann tensor, and therefore the metric perturbation, are functions of retarded time $u$  only. Therefore, $\partial_u p_{\alpha \beta} \neq 0$, where $\partial_u$ is the partial derivative with respect to retarded time. When written in terms of null coordinates, by virtue of the chain rule, only the partial derivative of the metric perturbation with respect to the $l^{\mu}$ tetrad is non-vanishing. Another important fact is that $q^{\alpha} = (q^l,0,0,0)$.

Using these facts, one can use tensor manipulations to show that a number of terms in Eqs.~\eqref{eq:FES2}-\eqref{eq:FET2} vanish identically. The reader familiar with these kind of manipulations should skip to below Eq.~\eqref{eq:FET2a}. 

\subsection*{Detailed manipulation of Equations~\eqref{eq:FES2}-\eqref{eq:FET2}}

We will analyze each term in Eqs.~\eqref{eq:FES2}-\eqref{eq:FET2} separately. Let us begin the analysis with Eq.~\eqref{eq:FES2}. Without loss of generality, we can use a coordinate system of the form $(u,v,x,y)$ where $u$ is retarded time and $v$ is advanced time, as defined in previous sections. The normal 4-vector is then of the form $n^\alpha = (1,0,0,0)$, as it points along the direction of propagation. Therefore, Eq.~\eqref{eq:FES2} takes the form,

\begin{align} \label{eq:FES2a}
\nabla^2 \Psi + \frac{ \alpha}{\kappa} \epsilon_{u u \delta \eta} n^u n^u q^u \vartheta (\partial^\eta {}^\delta {}_u \Psi)& \\ =   \nonumber \frac{\beta}{8 \kappa} q_\mu q^\mu \vartheta^2 + \frac{\beta}{4 \kappa} n^u n&^u q_u q_u  \vartheta^2 - \frac{\rho}{4 \kappa} 
\end{align}
\\
The second term on the left-hand side vanishes by definition of the Levi-Civita tensor, while the first term on the right-hand side vanishes because wave 4-vector is null, and the second term on the right-hand side vanishes because $q_u = 0$ [recall that $q^{u} \neq 0$, but $q_{u} = \eta_{uu} q^{u}$ and $\eta_{uu} = 0$, similar to what we have in Eq.~\eqref{eq:minkmat}]. Therefore, Eq.~\eqref{eq:FES2} is of the form,
\be \label{eq:FES2b}
\nabla^2 \Psi = - \frac{\rho}{4 \kappa}\,.
\ee

We will now continue to use a similar approach for Eq.~\eqref{eq:FEV2}. It is important to realize that $\Phi_{j}$ is transverse, which means it only has non-vanishing $x$ and $y$ components in the Cartesian basis or in our $(u,v,x,y)$ coordinate system. Let us begin by setting $j=x$ in Eq.~\eqref{eq:FEV2} (similar arguments would hold under the transformation $x \leftrightarrow y$), so that using the definitions of $n^\alpha$ and $q^\alpha$, Eq.~\eqref{eq:FEV2} reduces to
\allowdisplaybreaks[1]
\begin{widetext}
\begin{align} \label{eq:FEV2a}
\nonumber 2 \nabla^2 \Phi_j - \frac{4 \alpha}{\kappa} \epsilon_{x u y}{}^{v} q^u q^u \vartheta \partial_u {}_v \Phi^{y} + \frac{2 \alpha}{\kappa} \epsilon_{j u u \eta} n^u n^u q^u q^u \vartheta \partial_u {}^\eta \Phi_u  &- \frac{2 \alpha}{\kappa} \epsilon_{u u \delta \eta} n^u n^u q^u \vartheta \partial^\eta{}_u{}^{(\delta|} \Phi_{|j)} \\  - \frac{\alpha}{\kappa} \epsilon_{x u y }{}^v n^u n^u q^u \vartheta \partial_v{}_{u u} \Phi^{y} - \frac{2 \alpha}{\kappa} \epsilon_{u u \delta \eta} (n^u q_u) n^u q^u \vartheta \partial_j{}^\eta \Phi^\delta &= \frac{\beta}{2 \kappa} n^u q_u q_j \vartheta^2 - \frac{s_j}{2 \kappa} 
\end{align}
\end{widetext}
Since $\Phi_{j}$ is just a function of retarded time, it is clear that the second term and the fifth term on the left-hand side of Eq.~\eqref{eq:FEV2a} vanish. The third, fourth and sixth term on the left-hand side are zero by the properties of the Levi-Civita tensor, while the first term on the right-hand side is zero because $n^{\mu}$ points along $q^{\mu}$ and the latter is null by the equation of motion of the scalar field. We can thus see that Eq.~\eqref{eq:FEV2} is of the form
\be \label{eq:FEV2b}
\nabla^2 \Phi_{j} = - \frac{s_j}{4 \kappa}\,.
\ee

Equation~\eqref{eq:FES3} can be rewritten as,
\begin{align} \label{eq:FES3a}
\nonumber \delta_{jk} \nabla^2 (\Phi - \Psi) - \frac{\alpha}{2 \kappa}& \epsilon_{(k|u \beta \gamma} q^u \vartheta \partial^{\gamma \beta}{}_{|j)}(\Phi - \Psi) \\  =&~ - \frac{3}{4 \kappa} \delta_{jk} (\partial_t s + \tau) + \frac{\beta}{2 \kappa} q_j q_k \vartheta^2\,,
\end{align}
where here  $(j,k) \in (x,y)$. Since, $\Psi$ and $\Phi$ are functions of retarded time $u$ only, the second term on the left-hand side vanishes, since $\Psi_{,x} = 0 = \Psi_{,y}$ and $\Phi_{,x} = 0 = \Phi_{,y}$. The second term on the right-hand side also vanishes because the wave 4-vector $q^\mu$ points in the direction of propagation (along $u$). Therefore, with the above analysis, Eq.~\eqref{eq:FES3} can be expressed as 
\be \label{eq:FES3b}
\nabla^2(\Phi - \Psi) = - \frac{3}{4 \kappa} (\partial_t s + \tau)\,.
\ee

Let us finally look at Eq.~\eqref{eq:FET2}. The field $h^{TT}_{\alpha \beta}$ is the transverse-traceless part of the metric perturbation in our irreducible decomposition. The transverse nature of this term means that $\alpha$ and $\beta$ must be either $x$ or $y$ for a GW propagating along the $u$ direction. This implies that only the $(x,x)$, $(x,y)$ and $(y,y)$ components of $h^{TT}_{\alpha \beta}$ can be non-vanishing. This, in turn, implies that the fourth term on the left-hand side of Eq.~\eqref{eq:FET2} vanishes because $h^{TT}_{\alpha \beta}$ is contracted onto the wave vector $q^{\alpha}$ which points in the $u$ direction. Similarly, the second and the third terms of Eq.~\eqref{eq:FET2} can be shown to vanish using the arguments above and the dependence of $h^{TT}$ on retarded time. Therefore, we are left withs
\be \label{eq:FET2a}
\Box h^{TT}_{jk} = -\frac{\sigma_{jk}}{\kappa}\,.
\ee

\vspace{1cm}

We have now shown in excruciating detail that the complicated Eqs.~\eqref{eq:FES2}-\eqref{eq:FET2} reduce to Eqs.~\eqref{eq:FES2b},~\eqref{eq:FEV2b},~\eqref{eq:FES3b} and~\eqref{eq:FET2a}. The latter are exactly the same as Eq.~\eqref{FE1} of GR. Equations~\eqref{eq:FES2b},~\eqref{eq:FEV2b} and~\eqref{eq:FES3b} are (elliptic) Poisson equations, and therefore, at a particular time, the solutions depend on the matter configuration only, and the 4 degrees of freedom described by $\Psi$, $\Phi$ and $\Phi_{j}$ do not represent radiative modes. Equation~\eqref{eq:FET2a}, on the other hand, is a hyperbolic equation and it must thus represent a radiative mode corresponding to tensorial polarization modes $+$ and $\times$. These are the same GW modes that survive in GR, which confirms the results of Sec.~\ref{sec:NPdcs}.

%------------------------------------------------------------------------------------------------------------------------
\subsection{Irreducible decomposition in EdGB} \label{sec:irrededgb}

In this subsection, we present a brief calculation to obtain the polarization modes for EdGB gravity by decomposition of the metric into irreducible pieces. This subsection is very similar in essence to Sec. \ref{sec:irreddcs}. The decomposed equations of motion in EdGB gravity take the form

\begin{widetext}
\begin{align}
 2 \nabla^2 \Psi + 8 \lambda q^\beta q^\alpha \vartheta (\partial_\beta \partial_\alpha \Psi)  + 32 \lambda n^\alpha n^\beta q_\alpha q^\gamma \vartheta (\partial_\gamma \partial_\beta \Psi)& - 16 \lambda n^\alpha n^\beta q_\alpha q_\beta \vartheta (\nabla^2 \Psi) = - \rho + n^\alpha n^\beta q_\alpha q_\beta \vartheta^2 \\ \nonumber \\ \nonumber
  2 \nabla^2 \Phi_j - 16 \lambda n^\alpha n^\beta q^\gamma q_j \vartheta \partial_{\alpha \beta} \Phi_\gamma - 16 \lambda q^\alpha q^\beta& \vartheta \partial_{\alpha \beta} \Phi_j - 16 \lambda n^\alpha n^\beta q_\alpha q^\gamma \vartheta \partial_{\gamma \beta} \Phi_j\\ + 16 \lambda q^\alpha q^\beta \vartheta \partial_{j \beta} \Phi_\alpha  \nonumber + 16 \lambda n^\alpha n^\beta q_\alpha q^\gamma \vartheta \partial_{j \beta} \Phi_\gamma - 16 \lambda n^\alpha n^\beta q_\alpha q^\gamma \vartheta &\partial_{\gamma \beta} \Phi_j + 16 \lambda n^\alpha n^\beta q_\alpha q_j \vartheta \nabla^2 \Phi_\beta \\+ 16 \lambda n^\alpha n^\beta q_\alpha q_\beta \vartheta \nabla^2 \Phi_j &= -s_j + n^\alpha q_\alpha q_j \\ \nonumber \\
 \delta_{jk} \nabla^2 (\Phi - \Psi) =- \frac{3}{4 } \delta_{jk} &(\partial_t s + \tau) + q_j q_k \vartheta^2 \\ \nonumber
  \\ 
  \frac{1}{2} \Box h^{TT}_{jk}   + 8 \lambda q^\alpha q_{(k|} \vartheta \Box h^{TT}_{|j)\alpha} - 2 \lambda q^\alpha q^\beta \vartheta \partial_{(j|\beta} &h^{TT}_{|k)\alpha} - 4 \lambda \delta_{jk} q^\alpha q^\beta \vartheta \Box h^{TT}_{\alpha \beta} = - \sigma_{jk} 
\end{align}
\end{widetext}

Using the definitions of $q^\alpha$ and $n^\alpha$, along with the fact that the metric perturbation is only a function of retarded time, and noting that $\Phi_\alpha$ ($h^{TT}_{\alpha \beta}$) are transverse and thus they only possess non-vanishing $x$ and $y$ ($(x,x)$, $(x,y)$ and $(y,y)$) components, the above equations can be shown to reduce to the form 
\begin{align}\label{eq:S2} 
2 \nabla^2 \Psi + 8 \lambda q^u q^u \vartheta (\partial_u \partial_u \Psi)& = - \rho \,,  \\ 
\label{eq:V2} 2 \nabla^2 \Phi_j - 16 \lambda q^u q^u \vartheta \partial_{u u} \Phi_j &= -s_j \,, \\ 
\label{eq:S3} \delta_{jk} \nabla^2 (\Phi - \Psi) &= -\frac{3}{4} \delta_{jk} (\partial_t s + \tau) \,, \\  
\label{eq:T2} \frac{1}{2} \Box h^{TT}_{jk}   &= - \sigma_{jk} \,.
\end{align} 
after some tensor manipulations. However, recall that when we solved the wave equation for the scalar field [Eq.~\eqref{eq:CSFvacuum} ], we worked in the far field limit and kept only the leading $1/r$ term in the solution. Therefore, in the far field limit, the second term on left-hand side of Eqs.~\eqref{eq:S2} and~\eqref{eq:V2} are subdominant because they fall off a factor of $1/r$ faster than the first terms on the left-hand side of Eqs.~\eqref{eq:S2} and~\eqref{eq:V2}.

As in the dCS gravity case, the Eqs.~\eqref{eq:S2}-\eqref{eq:T2} of EdGB gravity are the same as those in GR [see Eq.~\eqref{FE1}], after discarding subdominant terms. Equations~\eqref{eq:S2}-\eqref{eq:S3} are elliptic, and thus, their solutions only depend on the matter configuration at a particular time instant, implying that $\Psi$, $\Phi$ and $\Phi_j$ are not radiative degrees of freedom. On the other hand, Eq.~\eqref{eq:T2} is hyperbolic, and thus, it describes a radiative degree of freedom. We can therefore conclude that, like GR, GWs in EdGB only possess two modes of polarization, namely the $+$ and $\times$ modes. This result is also in agreement with those obtained in Sec.~\ref{sec:NPedgb}.

%%%%%%%%%%%%%%%%%%%%%%%%%%%%%%%%%%%%%%%%%%%%%%
\section{Discussions} \label{sec:discuss}

We have here used two distinct methods to calculate and verify the polarization content of weak, plane, nearly null GWs in two different quadratic theories of gravity, namely dCS and EdGB gravity. The methods consisted of the Newman-Penrose formalism coupled to the E(2) classification, as well as an irreducible decomposition. We have found out that in both theories, the non-vanishing polarizations  are the two tensorial modes of GR (the $+$ and $\times$ GW modes).

This work, of course, is not the first to calculate the non-vanishing polarization modes in dCS and EdGB gravity. In the early 2000s, Jackiw and Pi \cite{Jackiw:2003pm} calculated the polarization modes for non-dynamical Chern Simons theory. This theory, however, is quite distinct from dCS gravity, because the scalar in the former was prescribed \emph{a priori} and not allowed to vary dynamically, rendering the theory overconstrained in certain scenarios~\cite{Sopuerta:2012de,Grumiller:2007rv}. The polarization modes in dCS gravity were studied later in \cite{Sopuerta:2012de}, considering a pp-wave spacetime and finding the same results obtained in our paper. The work of \cite{Sopuerta:2012de}, however, was limited to a pp-wave spacetime and only considered the evolution of the breathing mode, without studying the possibility of vector modes. Finally and more recently, \cite{SteinYunes} showed that as one approaches future null infinity $\mathscr{I}^+$, the trace of the metric perturbation obeys a wave equation, and thus, one can use the TT gauge to model GW polarizations in the far zone, implying that only the $+$ and $\times$ GW modes survive at $\mathscr{I}^+$. The work of \cite{SteinYunes}, however, did not study why or how the other potential polarization modes are suppressed in dCS gravity.

The results presented here confirm and extend the results of earlier papers on dCS gravity, arriving at the same conclusions by exploring the why and the how in more detail through two techniques that had not been explored before. Moreover, we applied the same techniques to EdGB gravity, arriving again to the same results as in dCS gravity, though it seems this is the first time these results appear in the literature. Our work therefore shows, in a pedagogical way, how to calculate the evolution equation for different polarization modes in modified theories of gravity. 

Our results also have important implications for gravitational wave tests of GR. In the near future, the detection of GWs through multiple interferometers, or through space-based instruments, hold the key to measure the polarization content of GWs. If future observations can show that only the two $+$ and $\times$ modes are present in nature, this could be a death-blow to many modified theories. We here show clearly that this is not the case in general. In dCS and EdGB gravity, as well as probably in other theories of gravity, the polarization content of GWs remains the same as in GR, and thus, polarization tests of GR with GWs are uninformative. The best avenue to constrain these theories, therefore, continues to be the dynamical late inspiral and merger phase of coalescing binaries~\cite{Alexander:2018qzg}. 

%%%%%%%%%%%%%%%%%%%%%%%%%%%%%%%%%%%%%%%%%%%%%%
\section{Acknowledgements}

We would like to thank David Garfinkle and Leo Stein for useful discussions. Nicol\'as Yunes acknowledges support from NSF grant PHY-1759615, NASA grants 80NSSC18K1352. 

%%%%%%%%%%%%%%%%%%%%%%%%%%%%%%%%%%%%%%%%%%%%%%
\appendix
\section{Detailed calculation of the polarization modes in EdGB gravity using the NP method} \label{Appendix A}

Let us begin by considering the scalar field evolution controlled by Eq. (\ref{eq:EomEdgb2}). Since the Gauss-Bonnet invariant $\mathcal{G}$, defined by Eq. (\ref{eq:edgbscalar}), decays fast in the far zone, the interaction term vanishes. One is then left with a free wave equation in flat spacetime, whose solution is simply
\be \label{eq:wavesoln}
\vartheta = B e^{i q^\mu x_\mu} \,.
\ee
with $B$ an amplitude and $q^\mu$ the 4-wave vector, which is also null, i.e.~$ q^\mu q_\mu = 0$. 

Let us now consider the trace-reversed form of the field equations given in Eq. (\ref{eq:tricci3}). Using Eq. (\ref{eq:wavesoln}) in Eq. (\ref{eq:tricci3}), we have
\begin{align} \label{eq:tricciedgb}
\nonumber R_{\mu \nu} =&  - B^2 q_\mu q_\nu e^{2i q\cdot x} \\&~~~~+ 2 \lambda~B~ \delta^{\gamma \delta \kappa \epsilon}_{\alpha \beta \rho \sigma} R^{\rho \sigma}{}_{\kappa \epsilon} q^\alpha q_\gamma e^{i q \cdot x}\delta^\beta {}_{(\mu} \eta_{\nu ) \delta} \,.
\end{align}
Since we are working with vacuum, the first term in the equation above vanishes. With this at hand, we can now compute the different NP scalars. 

%------------------------------------------------------------------------------------------
\subsubsection{Analysis for $\Psi_{2}$}

Equation (\ref{eq:tricciedgb}) implies that
\begin{align} \label{eq:psi21}
\nonumber \Psi_{2}  = - \frac{1}{6} R_{nl} =&  \frac{1}{6} B^2 q_n q_l e^{2i q\cdot x} \\&~~~~
- \frac{1}{3} \lambda~B~ \delta^{\gamma \delta \kappa \epsilon}_{\alpha \beta \rho \sigma} R^{\rho \sigma}{}_{\kappa \epsilon} q^\alpha q_\gamma e^{i q \cdot x}\delta^\beta {}_{(n} \eta_{l ) \delta} \,.
\end{align}
For a weak, plane, nearly null GW, the Riemann tensor is only dependent on retarded time. Thus, Eq. (\ref{eq:EomEdgb2}) along with Eq. (\ref{eq:wavesoln}) imply that the only non-vanishing component is along retarded time, i.e.~either $q^l$ (or $q_{n}$ when using the metric in Eq. (\ref{eq:minkmat})) are the only non-vanishing components. Using this, the first term of Eq. (\ref{eq:psi21}) vanishes and one is left with
\begin{equation} \label{eq:psi22}
 \Psi_{2} = -\frac{1}{3} \lambda~B~ \delta^{\gamma \delta \kappa \epsilon}_{\alpha \beta \rho \sigma} R^{\rho \sigma}{}_{\kappa \epsilon} q^\alpha q_\gamma e^{i q \cdot x}\delta^\beta {}_{(n} \eta_{l ) \delta} \,.
\end{equation} 
Using Eq. (\ref{eq:minkmat}) and the non-vanishing components of the wave 4-vector, we see that Eq. (\ref{eq:psi22}) vanishes due to the symmetries of the generalized Kronecker delta. The conclusion then is that $R_{nl} = 0$, and thus, $\Psi_{2} =0$.

%------------------------------------------------------------------------------------------
\subsubsection{Analysis for $\Psi_{3}$} 

Equation (\ref{pol2}) implies that
\begin{align} \label{eq:psi31}
\nonumber \Psi_{3}  = - \frac{1}{2} R_{n \bar{m}} =&  \frac{1}{2} B^2 q_n q_{\bar{m}} e^{2i q\cdot x} \\&~~~~ -  \lambda~B~ \delta^{\gamma \delta \kappa \epsilon}_{\alpha \beta \rho \sigma} R^{\rho \sigma}{}_{\kappa \epsilon} q^\alpha q_\gamma e^{i q \cdot x}\delta^\beta {}_{(n} \eta_{\bar{m} ) \delta} \,.
\end{align}
The first term in the above equation vanishes again, since $q_n$ is the only non-zero component of the 4-wave vector for the GW under consideration. Using Eq. (\ref{eq:minkmat}), Eq. (\ref{eq:psi31}) can be written as 
\begin{align} \label{eq:psi32}
\nonumber \Psi_{3}  =&  \frac{1}{2} \lambda~B~ \delta^{n m \kappa \epsilon}_{l n \rho \sigma} R^{\rho \sigma}{}_{\kappa \epsilon} q^l q_n e^{i q \cdot x} \\ &~~~~~~ - \frac{1}{2}  \lambda~B~ \delta^{n l \kappa \epsilon}_{l \bar{m} \rho \sigma} R^{\rho \sigma}{}_{\kappa \epsilon} q^l q_n e^{i q \cdot x} \,.
\end{align}
The properties of the generalized Kronecker delta restrict the values that the indices $\rho$, $\kappa$, $\epsilon$, and $\sigma$ can take. We can therefore use Eq. (\ref{linriem1}) to rewrite the Riemann tensor in Eq. (\ref{eq:psi32}). Since the metric or the metric perturbation is only a function of retarded time, we find that the right hand side of Eq. (\ref{eq:psi32}) is zero identically. Therefore, with Eq. (\ref{pol2}), we have that
\be
\Psi_{3} =0\,.	
\ee

%------------------------------------------------------------------------------------------
\subsubsection{Analysis for $\Phi_{22}$}

Equation (\ref{eq:tricciedgb}) implies that
\begin{align} \label{eq:phi1}
\nonumber \Phi_{22} &= - \frac{1}{2}  R_{n n} =  \frac{1}{2} B^2 q_n q_n e^{2i q\cdot x} \\&~~~~ -\lambda~B~ \delta^{\gamma \delta \kappa \epsilon}_{\alpha \beta \rho \sigma} R^{\rho \sigma}{}_{\kappa \epsilon} q^\alpha q_\gamma e^{i q \cdot x}\delta^\beta {}_{(n} \eta_{n) \delta} \,.
\end{align}
The second term in the above equation vanishes by the same argument presented in the paragraph below Eq. (\ref{eq:psi32}). The first term does not vanish, but it scales as $\mathcal{O}(r^{-2})$ in the far zone. Therefore, we have that far from the source
\be \label{eq:phi2}
\Phi_{22} = 0\,.
\ee

%------------------------------------------------------------------------------------------
\subsubsection{Analysis of $\Psi_4$}

Using the same line of reasoning, there are no constraints one can place on $\Psi_4$ using the field equations for EdGB gravity. This in turn means that, just like in dCS gravity, $\Psi_4$ remains unconstrained or non-vanishing in EdGB gravity. Therefore, the transverse-traceless tensorial modes, i.e.~the $+$ and $\times$ GW polarizations, are non-vanishing in EdGB gravity for a weak, plane, nearly null GW.

\bibliography{ppr_v}	
	
\end{document}